\documentclass[showpacs,amsmath,amssymb,aps,showkeys,floatfix,prd,a4paper]{revtex4}

\usepackage[dvips]{graphicx}
\usepackage{dcolumn}
\usepackage{bm}
\usepackage{epsfig}
\usepackage{amsfonts}
\usepackage{amssymb,amscd}
\usepackage{subfigure}

\newcommand{\rx}{\mbox{\boldmath $x$}}
\newcommand{\ry}{\mbox{\boldmath $y$}}

\def\lsim{\raise0.3ex\hbox{$<$\kern-0.75em\raise-1.1ex\hbox{$\sim$}}}

\def\gsim{\raise0.3ex\hbox{$>$\kern-0.75em\raise-1.1ex\hbox{$\sim$}}}

\newcommand{\be}{\begin{equation}}

\newcommand{\ee}{\end{equation}}

\def\beq{\begin{equation}}

\def\eeq{\end{equation}}

\def\beqa{\begin{eqnarray}}

\def\eeqa{\end{eqnarray}}

\newcommand{\ba}{\begin{eqnarray}}

\newcommand{\rr}{\mbox{\boldmath $r$}}

\newcommand{\rb}{\mbox{\boldmath $b$}}

\def\gappeq{\mathrel{\rlap {\raise.5ex\hbox{$>$}}

{\lower.5ex\hbox{$\sim$}}}}

\def\lappeq{\mathrel{\rlap{\raise.5ex\hbox{$<$}}

{\lower.5ex\hbox{$\sim$}}}}

\def\Toprel#1\over#2{\mathrel{\mathop{#2}\limits^{#1}}}

\begin{document}

\title{Gluon saturation and Feynman scaling in leading neutron production}
\author{F. Carvalho$^{1}$, V.P. Gon\c{c}alves$^{2}$,  D. Spiering$^{3}$  and  F.S. Navarra$^3$}
\affiliation{$^1$Departamento de Ci\^encias Exatas e da Terra, Universidade Federal de S\~ao Paulo,\\  
Campus Diadema, Rua Prof. Artur Riedel, 275, Jd. Eldorado, 09972-270, Diadema, SP, Brazil.\\ 
$^{2}$High and Medium Energy Group, Instituto de F\'{\i}sica e Matem\'atica,  Universidade Federal de Pelotas\\
Caixa Postal 354,  96010-900, Pelotas, RS, Brazil.\\
$^3$Instituto de F\'{\i}sica, Universidade de S\~{a}o Paulo,
C.P. 66318,  05315-970 S\~{a}o Paulo, SP, Brazil.\\
}

\begin{abstract}
In this paper we extend the color dipole formalism for the study of leading neutron production in  $e + p \rightarrow e + n + X$ collisions 
at high energies and estimate the related  observables which were measured  at HERA and could be analysed in future electron-proton ($ep$) 
colliders. In particular, we calculate the Feynman $x_F$ distribution of leading neutrons, which is expressed in terms of the pion flux and 
the photon-pion total cross section.  In the color dipole formalism, the photon-pion cross section is described in terms of the dipole-pion scattering amplitude,   which contains information about the QCD dynamics at high energies  and gluon saturation effects. 
We consider different models for the scattering amplitude, which have been used to describe the inclusive and diffractive $ep$ HERA data. 
Moreover, the model dependence of our predictions with the description of the pion flux is analysed in detail. We demonstrate the  recently 
released H1 leading neutron spectra can be described using the color dipole formalism and that  these spectra could help us to observe more 
clearly gluon saturation effects  in future $ep$ colliders.

\end{abstract}



\maketitle

\vspace{1cm}

\section{Introduction}

In high energy collisions the outgoing baryons which have large fractional longitudinal momentum ($x_L \geq 0.3$) and the same  valence quarks 
(or at least one of them) as the incoming particles are called leading particles (LP).   
The momentum spectra of leading particles have been measured already some time ago \cite{lpdata}.  
Very recently,  high precision data on leading  neutrons produced in electron-proton reactions at HERA at high energies     
became available \cite{lpdata2}.  

\begin{figure}
\centerline{\psfig{figure=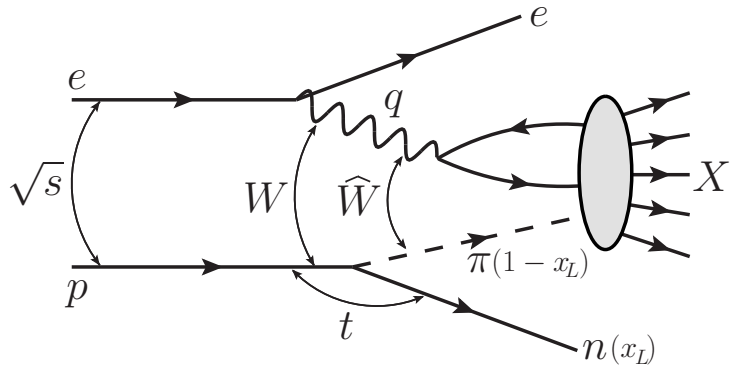,width=10cm}}  
\caption{Leading neutron $n$ production in  $e p \rightarrow e n X $ interactions at high energies. 
$x_L$ is momentum fraction of the proton carried by the neutron $n$.}
\label{fig1}
\end{figure}

Leading neutron production   is a very interesting process. In spite of more than ten years of 
intense experimental and theoretical efforts, the  $x_L$ (Feynman momentum) distribution of leading neutrons remains 
without a satisfactory theoretical description 
\cite{holt,bisha,kope,kuma,niko99,models,kkmr,khoze,speth,pirner}. Monte Carlo studies, using standard deep inelastic 
scattering (DIS) generators show \cite{lpdata2,monte} that these processes have a rate of neutron production a factor of three lower than the data and produce a 
neutron energy  spectrum with the wrong shape, peaking at values of $x_L$ below $0.3$.  
In order to fit the data the existing models need to combine  different ingredients including pion exchange, reggeon exchange, baryon resonance excitation 
and decay and independent fragmentation \cite{models,kkmr,khoze,speth,pirner}.  Moreover, the incoming photon interacts with the pion emitted by proton and then 
rescatters, interacting  
also with the emerging neutron and giving origin to significant  absorptive corrections, which are difficult to calculate \cite{speth,pirner}.
As it can be seen in Fig. \ref{fig1}, this process is essentially composed by a soft pion (or reggeon) emission and by the 
subsequent photon-pion interaction at high energies. Pion emission has been studied for a long time and according to 
the traditional wisdom   it can be described by a simple interaction Lagrangian of the form 
$g \bar{\psi} \gamma_5 \pi \psi$, where $\psi$ and $\pi$ are the nucleon and pion field respectively. 
The corresponding pion-nucleon amplitude must be supplemented with a form factor, which 
represents the extended nature of hadrons and at the same time regularizes divergent integrals. The precise functional 
form of the form factor is very hard (if not impossible) to extract from first principle calculations. We must then 
resort to models. Very recently \cite{wally,wally2} it has been argued that the Lagrangian which is  more 
consistent with the chiral symmetry  requirements is the one with a pseudovector couplings.    
One of the goals of the present study is to test in a new context the 
better founded splitting function, $f(y)$, derived in \cite{wally}. Additionally, forward hadron production is very important also for cosmic ray physics, 
where highly energetic protons reach the top of the atmosphere and undergo successive high energy scatterings off  light nuclei in the air. In each of 
these  collisions, a  projectile proton  
(the leading  baryon) looses energy, creating showers of particles, and goes to the next scattering.  The interpretation of cosmic  data depends on the accurate 
knowledge of the leading baryon momentum spectrum  and its energy dependence. 
A crucial question of practical importance is the existence or non-existence of the Feynman scaling, which says 
that the $x_L$ spectra  of secondaries are energy independent. In cosmic ray applications we are sensitive essentially 
to the large $x_L$ region (the fragmentation  region), which probes the   low Bjorken-$x$ component of the target 
wave function. In this kinematical range nonlinear effects are expected to be present in the description of the QCD dynamics  
(for recent reviews see \cite{cgc}), associated to the high parton density. The state-of-art framework to treat  QCD at high energies is 
the Color Glass Condensate (CGC) formalism \cite{CGC2}, which predicts  gluon saturation at small-$x$, with the evolution with the energy being 
described by an  infinite hierarchy of coupled equations for the correlators of  Wilson lines -- the Balitsky-JIMWLK  hierarchy \cite{cgc}.  
In the mean field approximation, this set of equations can be approximated by the Balitsky-Kovchegov (BK) equation \cite{BAL,kov}, which describes the 
evolution of the dipole-target scattering amplitude with the rapidity $Y = \ln(1/x)$.

In this paper we propose to treat the leading neutron production at HERA using the color dipole formalism, which is able to describe the  inclusive and diffractive HERA $ep$ data. Our goal is to extend this successful formalism, which have its main parameters well determined, for leading neutron production. As a consequence, our predictions are free parameter, depending only from the choices for the models of the pion flux and dipole scattering amplitude.  Moreover, the use of the color dipole formalism allows to estimate the contribution of the gluon saturation effects for the leading neutron production in the kinematical range which was probed by HERA and which will probed in future electron-proton colliders.  Finally, it allows to investigate the relation between the Feynman scaling (or its violation) and the description of the QCD dynamics at high energies. It is important to emphasize that, in the near future, Feynman scaling will be investigated  experimentally at the LHC by the LHCf Collaboration \cite{lhcf,lhcf2,lhcf3}. 

This paper is organized as follows. In the next Section we present a brief review of the leading neutron production in $ep$ collisions as well as the different models for the pion flux are discussed. Moreover, the treatment of the process using the color dipole formalism is presented and the main assumptions are analysed. In Section \ref{results} we analyse the dependence of our predictions in the pion flux and in the scattering amplitude. A comparison with the recent H1 data is presented and the Feynman scaling is analysed. Finally, in Section  \ref{conc} we summarize our main conclusions.

\section{Leading neutron production in the color dipole formalism}
\label{formalismo}

\subsection{The cross section}

Let us review the main formulas of leading neutron production. At high energies, this scattering  can be seen 
as a set of three factorizable subprocesses (See Fig. \ref{fig1}):  i) the photon fluctuates into a 
quark-antiquark pair (the color dipole), ii) the color dipole interacts with the pion, present in the wave function of the incident 
proton, and iii) the leading neutron is formed. The differential cross section reads: 
\beq
\frac{d^2 \sigma(W,Q^2,x_L,t)}{d x_L d t} = f_{\pi/p} (x_L,t) \sigma_{\gamma^* \pi}(\hat{W}^2,Q^2)
\label{crossgen}
\eeq
where $Q^2$ is the virtuality of the exchanged photon, $\hat{W}$ is the center-of-mass energy of the 
virtual photon-pion system. It can be written as $\hat{W}^2 = (1-x_L) \, W^2$, where $W$ is the center-of-mass energy of the 
virtual photon-proton system. As it can be seen in Fig. \ref{fig1}, $x_L$ is the proton momentum fraction carried by the 
neutron and $t$ is the square of the four-momentum of the exchanged pion. In terms of the measured quantities $x_L$ and 
transverse momentum $p_T$, the pion 
virtuality is:
\beq
t \simeq-\frac{p_T^2}{x_L}-\frac{(1-x_L)(m_n^2-m_p^2 x_L)}{x_L}
\label{virtuality}
\eeq
The flux of virtual pions emitted by the proton is represented by $f_{\pi/p}$ and $\sigma_{\gamma^* \pi}(\hat{W}^2,Q^2)$  
is the cross section for the interaction between the  virtual-photon and the virtual-pion  at center-of-mass energy $\hat{W}$.

\subsection{The pion flux}

The  pion flux $f_{\pi/p} (x_L,t)$ (also called sometimes pion splitting function) is the virtual pion momentum distribution
in a physical nucleon (the bare nucleon plus the ``pion cloud''). It was first calculated in the early studies 
\cite{sullivan} of deep inelastic 
scattering (DIS), where a pseudoscalar nucleon-pion-nucleon vertex was added to the standard DIS diagram. These early calculations 
were further refined in Refs. \cite{cloud} and even extended to the strange and charm sector \cite{charm}.  
Since pion emission is a nonperturbative 
process, once we depart from the single nucleon state, we should consider a whole tower of meson-baryon states, having to deal 
with a series for which there is no rigorous truncation scheme. This led some authors to use light 
cone models \cite{ma} 
for the pion momentum distribution, where the dynamical  origin of the pion is not mentioned and one tries to determine 
phenomenologically the relative weight of the higher Fock states.

In all the calculations of the pion flux a form factor was introduced to represent the non pointlike nature of
hadrons and hadronic vertices, which  contain a cut-off parameter determined by fitting data.
The most frequently used parametrizations of the pion flux  
\cite{holt,bisha,kope,kuma,niko99,models,kkmr,khoze,speth,pirner} have the  
following  general form:
\beq
f_{\pi/p} (x_L,t)  = \frac{1}{4 \pi} \frac{2 g_{p \pi p}^2}{4  \pi} \frac{-t}{(t-m_{\pi}^2)^2} (1-x_L)^{1-2 \alpha(t)}  
[F(x_L,t)]^2
\label{genflux}
\eeq 
where $g_{p \pi p}^2/(4 \pi) = 14.4$ is the $ \pi^ 0 p p $ coupling constant, $m_{\pi}$ is the pion mass and $\alpha(t)$ will be 
defined below. The form factor $F(x_L,t)$  accounts for the finite size of the nucleon and pion. We will consider the following parametrizations 
of the form factor:
\beq
F_1(x_L,t) = \exp \left[ R^2 \frac{(t-m_{\pi}^2)}{(1-x_L)} \right] \,\,\,\, , \,\,\,\, \alpha(t) = 0
\label{form1}
\eeq
from Ref. \cite{holt}, where $R = 0.6$ GeV$^{-1}$. 
\beq
F_2(x_L,t) = 1 \,\,\,\, , \,\,\,\, \alpha(t) = \alpha(t)_{\pi}
\label{form2}
\eeq
from  Ref. \cite{bisha}, where $\alpha_{\pi}(t) \simeq t$ (with $t$ in GeV$^2$) is the Regge trajectory of the pion.
\beq
F_3(x_L,t) =  \exp \left[ b (t-m_{\pi}^2) \right] \,\,\,\, , \,\,\,\, \alpha(t) = \alpha(t)_{\pi} 
\label{form3}
\eeq
from Ref. \cite{kope}, where $b = 0.3$ GeV$^{-2}$. 
\beq
F_4(x_L,t) =  \frac{\Lambda_m^2-m_{\pi}^2}{\Lambda_m^2-t}      \,\,\,\, , \,\,\,\, \alpha(t) = 0
\label{form4}
\eeq
from Ref. \cite{kuma}, where $\Lambda_m = 0.74$ GeV. 
\beq
F_5(x_L,t) =  \left[\frac{\Lambda_d^2-m_{\pi}^2}{\Lambda_d^2-t}\right]^2      \,\,\,\, , \,\,\,\, \alpha(t) = 0
\label{form5}
\eeq
also from Ref. \cite{kuma}, where $\Lambda_d = 1.2$ GeV. In the case of the more familiar exponential (\ref{form1}), monopole (\ref{form4})  and dipole (\ref{form5}) 
forms factors, the cut-off parameters have been determined by fitting low energy data on nucleon and nuclear reactions and also data on deep inelastic scattering 
and structure functions.

In all  analyses of pion cloud effects in DIS the authors  have used pseudoscalar (PS)
coupling, which is, by itself, inconsistent with chiral symmetry.  
More recently, in \cite{wally} the authors used an  effective chiral Lagrangian for the interaction of pions and
nucleons consistent with chiral symmetry. Unlike the earlier chiral effective
theory calculations which only computed the light-cone
distributions of pions  or considered the non analytic
behaviour of their moments to lowest order in the pion mass, in \cite{wally} the authors have computed the complete set 
of diagrams relevant for DIS from nucleons dressed by pions resulting
from their Lagrangians, without taking the heavy
baryon limit. In particular, they have demonstrated explicitly
the consistency of the computed distribution functions
with electromagnetic gauge invariance.  In \cite{wally2} they applied the previously computed pion distribution to the study 
of the $\bar{d}-\bar{u}$ asymmetry in the nucleon.  For our purposes, the  relevant contribution of the pion momentum distribution
reads \cite{wally}:  
\beq
f_{\pi/p} (y)  = \frac{ g_{A}^2 m_n^2}{(4 \pi f_{\pi})^2 } \int_0^{\Lambda_c^2} d p_T^2 
\frac{y(p_T^2 + y^2 m_n^2)}{[p_T^2 + y^2 m_n^2 + (1-y)m_{\pi}^2]^2} 
\label{form6}
\eeq
where $y=1-x_L$, $g_A = 1.267$ is the nucleon axial charge, $f_{\pi} = 93$ MeV 
and $\Lambda_c = 0.2$ GeV. In order to obtain the final cross section we multiply 
(\ref{form6}) by 2 (see \cite{wally,wally2}) and insert it into  (\ref{crossgen}) after 
integrating the latter over $t$ (or over $p_T$).

In what follows, we shall use the  pion fluxes listed above,  in equations  (\ref{form1})-(\ref{form6}), denoting them by   
$f_1$, $f_2$,  ... $f_6$, respectively.  The main purpose of our calculation will be to show that the dipole approach gives a good description 
of data and the pion flux is just one element of the calculation. Therefore we will make no effort to choose one particular form. Nevertheless, we note 
that one important phenomenological constraint that these pion fluxes must satisfy is to reproduce the $\bar{d}-\bar{u}$ asymmetry in the proton sea 
measured by the E866 Collaboration \cite{e866}. Among the pion fluxes mentioned above only (\ref{form5}) (see \cite{babi00}) and (\ref{form6}) have been 
confronted with these  data.  As  it will be seen later, the results discussed here are very sensitive to the choice of the pion flux and hence  leading 
neutron spectra may be used to constrain its shape.

Before closing this subsection we would like to mention that the diagram shown in Fig.\ref{fig1} represents only the dominant contribution to 
leading neutron production. Other isovector  meson exchanges, such as $\rho$ or $a_2$, can also contribute to the leading neutron spectrum.  Moreover 
the $p \rightarrow \Delta$ transition can also contribute to neutron production through the subsequent decay $\Delta \rightarrow n \pi$. 
Theoretical studies 
show that processes other than direct pion exchange are expected to give a contribution of less than $25$ \% of the cross section 
\cite{holt,kope,kuma,niko99,iso}. The effect of these backgrounds to the one pion exchange is to increase the rate of neutron production. 
However this effect is partially compensated by the absorptive 
rescattering of the neutron, which decreases the neutron rate by  approximately the same amount \cite{speth,pirner}. 

\subsection{The photon-pion cross section}

In order to obtain the photon-pion cross section we will use the  color dipole formalism, as  usually done in high energy deep inelastic scattering 
off a nucleon target. In this formalism,  the cross section is factorized in terms of the photon wave functions $\Psi$, which describes the photon splitting in a $q\bar{q}$ pair, and the dipole-pion cross section  $\sigma_{d\pi}$. We have 
\begin{equation} 
\sigma_{\gamma^* \pi} (\hat{x}, Q^2)   
=   \int _0 ^1 dz \int d^2 \rr \sum_{L,T} \left|\Psi_{T,L} (z, \rr, Q^2)\right|^2  \sigma_{d\pi}(\hat{x}, \rr)
\label{f3}   
\end{equation}
where
\beq
\hat{x} = \frac{Q^2 + m^2_f}{\hat{W}^2 + Q^2} = \frac{Q^2 + m^2_f}{(1-x_L)W^2 + Q^2}
\label{xhat}
\eeq
is the scaled Bjorken variable and the variable $\rr$ defines the relative transverse separation of the pair (dipole). Moreover, we have that the photon wave functions are given by 
\begin{eqnarray}
\label{eq:psi^2}
\left|\psi_{L}(z,\rr)\right|^{2} & =  & \frac{3 \alpha_{em}}{2 \pi^{2}}\sum_f e_{f}^
{2} 4 Q^{2}z^{2}(1-z)^{2} K_{0}^{2}(\epsilon r) \\
  \left|\psi_{T}(z,\rr)\right|^{2} & = & \frac{3 \alpha_{em}}{2 \pi^{2}}\sum_{f} e_{f}^
{2} \left\{[z^{2} + (1-z)^{2}] \epsilon^{2} K_{1}^{2}(\epsilon r) + m_{f}^{2} 
K_{0}^{2}(\epsilon r) \right\} 
\end{eqnarray}
for a longitudinally (L) and transversely (T)  polarized photon,  respectively. In the above 
expressions  \( \epsilon^{2} = z (1- z)Q^{2} + m_{f}^{2}\; ,\) 
$K_{0}$ and $K_{1}$ are modified Bessel functions and the sum is over quarks of flavour 
$f$  with  a corresponding  quark mass $m_f$. As usual $z$ stands for  the 
longitudinal photon momentum fraction carried by the quark and $1-z$ is the 
longitudinal photon momentum fraction of  the antiquark.   

The main input in the calculations of $\sigma_{\gamma^* \pi}$ is  the 
dipole-pion cross section. In what follows, for simplicity,  we will assume the validity of 
the additive quark model, 
which allows us to relate $\sigma_{d\pi}$ with the dipole-proton cross section, usually probed 
in the typical inclusive and exclusive processes at HERA. Basically, we will assume that
\begin{equation}
\sigma_{d\pi} ({x}, \rr) = \frac{2}{3} \cdot \sigma_{dp} ({x}, \rr) 
\label{doister}
\end{equation}
This is assumption is supported by the study of the pion structure function in the low x
regime presented in \cite{zoller}. It also gives a good description of the previous ZEUS 
leading neutron spectra, as shown in \cite{kkmr,khoze}. On the other hand,  
the direct application of  (\ref{crossgen}) to HERA photoproduction 
data \cite{zeus02}  leads to the result 
$ \sigma^{\mbox{tot}}_{\gamma \pi} / \sigma^{\mbox{tot}}_{\gamma p} = 0.32 \pm 0.03$, 
which is factor 2 lower than the  ratio given above.  This subject certainly deserves 
more detailed studies in the future. For our purposes, the use of  relation (\ref{doister})  
allows us to estimate $\sigma_{d\pi}$ without additional 
free parameters.  In the eikonal approximation the  
dipole-proton cross section $\sigma_{dp}$  is given by:
\begin{equation} 
\sigma_{dp} (x, \rr) = 2 \int d^2 \rb \,  {\cal N}^p (x, \rr, \rb)\,\,,
\label{sdip}
\end{equation}
where  $\mathcal{N}^p(x,\rr,\rb)$ is  the imaginary part of the forward amplitude for the scattering between a small dipole
(a colorless quark-antiquark pair) and a dense hadron target, at a given
rapidity interval $Y=\ln(1/x)$. The dipole has transverse size given by the vector
$\rr=\rx-\ry$, where $\rx$ and $\ry$ are the transverse vectors for the quark
and antiquark, respectively, and impact parameter $\rb=(\rx+\ry)/2$. 
As mentioned in the introduction, at very high energies (and very low $x$) the evolution with the rapidity $Y$ of  $\mathcal{N}^p(x,\rr,\rb)$  
is given  by the Balitsky-Kovchegov (BK) equation \cite{BAL,kov} assuming the  translational invariance approximation, which implies  $\mathcal{N}^p(x,\rr,\rb) = \mathcal{N}^p(x,\rr) S(\rb)$ and $\sigma_{dp} (x, \rr) = \sigma_0 \cdot  {\cal N}^p (x, \rr)$, with the normalization of the dipole cross section ($\sigma_0$) being fitted to data.
Alternatively, we can  describe the scattering amplitude $\mathcal{N}^p(x,\rr)$ using phenomenological models based on saturation physics constructed taking into account the analytical solutions of the BK equation which are known in the low and high density regimes. The main advantage in the use of phenomenological models is that we can easily to compare the linear and nonlinear predictions, which is useful to determine the contribution of the saturation effects for the process under analysis. 
In what follows we will  consider as input the phenomenological models proposed in Refs. \cite{GBW, iim,soyez}. 
In particular, the IIMS model, proposed in Ref. \cite{iim}
and updated in \cite{soyez},  was  constructed so as 
to reproduce two limits  of the LO BK equation 
analytically under control: the solution of the BFKL equation
for small dipole sizes, $r\ll 1/Q_s(x)$, and the Levin-Tuchin law 
for larger ones, $r\gg 1/Q_s(x)$. In the updated version of this parametrization \cite{soyez}, the free parameters were obtained by 
fitting the new H1 and ZEUS data.
In this parametrization the  forward dipole-proton scattering amplitude is given by
\begin{eqnarray}
{\cal{N}}^p(x,\rr) =  \left\{ \begin{array}{ll} 
{\mathcal N}_0\, \left(\frac{r\, Q_s}{2}\right)^{2\left(\gamma_s + 
\frac{\ln (2/r\, Q_s)}{\kappa \,\lambda \, Y}\right)}\,, & \mbox{for $r 
Q_s({x}) \le 2$}\,,\\
 1-\text{e}^{-a\,\ln^2\,(b\,r\, Q_s)}\,,  & \mbox{for $r Q_s({x})  > 2$}\,, 
\end{array} \right.
\label{CGCfit}
\end{eqnarray}
where  $a$ and $b$ are determined by continuity conditions at $r Q_s({x})=2$,  
$\gamma_s= 0.7376$, $\kappa= 9.9$, ${\mathcal N}_0=0.7$
and $Q_s$ is the saturation scale given by: 
\beq
Q^2_s ({x}) = Q^2_0 \left( \frac{x_0}{{x}}\right)^{\lambda}
\label{qsat}
\eeq
with $x_0=1.632\times 10^{-5}$, $\lambda=0.2197$, $Q_0^2 = 1.0$ GeV$^2$.
The first line of Eq. (\ref{CGCfit}) describes the linear regime whereas the second one includes saturation effects. 
On the other hand, in the GBW model \cite{GBW}, the dipole-proton scattering amplitude is given by
\beq
{\cal{N}}^p(x,\rr) =  1- \exp \left[ -\frac{ Q_s^2 r ^2}{4}  \right]
\label{gbw}
\eeq
with $Q_0^2 = 1.0$ GeV$^2$,  $x_{0}=3\times 10^{-4}$ and $\lambda=0.288$.
Finally, we will also use the dipole-proton cross section estimated with the help of the DGLAP analysis of the gluon distribution, which is 
given by \cite{predazzi}:
\beq
\sigma_{dip}(x,\rr) = \frac{\pi^2}{3}r^2\alpha_s {x}g(x,10/r^2)
\label{sig_dglap}
\eeq
where $x g(x,Q^2)$ is the target gluon distribution, for which we use the CTEQ6 parametrization \cite{lai}.  The above expression represents the linear regime 
of QCD and is a baseline for comparison with the nonlinear predictions.

\section{Results}
\label{results}

\begin{figure}[t]
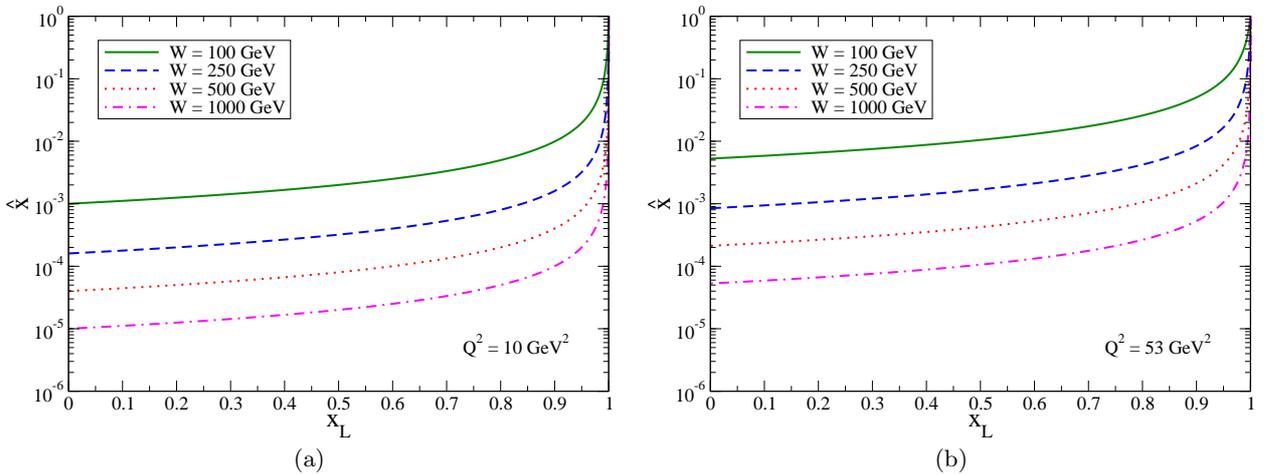

\begin{tabular}{ccc} 
\psfig{figure=leadf2an.eps,width=8cm}& \,\,\, & \psfig{figure=leadf2bn.eps,width=8cm}\\
(a) & \,\,\, &(b)
\end{tabular}
\caption{Scaled Bjorken variable $\hat{x}$, (Eq. \ref{xhat}),   as a function of $x_L$ for two different virtualities: 
(a)  $Q^2=10$ GeV$^2$ and  
(b)  $Q^2=53$ GeV$^2$.}
\label{fig2}
\end{figure}

As it was seen in the previous sections, the leading neutron spectrum depends essentially on two main ingredients: the pion flux and the dipole-pion cross 
section. The latter is very sensitive to the value of the involved Bjorken $x$, or, in our case, $\hat{x}$. In Fig. \ref{fig2}  we show  $\hat{x}$ as a 
function of $x_L$.  From the figure we can conclude that at the highest values of the present photon-hadron energies ($W$) and at the lowest values of $Q^2$ we enter 
deeply in the low $x$ domain. This will be even more so if  measurements can be carried out 
at higher energies, but already at the available energies we can see that the leading neutron spectrum receives a contribution from the kinematical range where the  nonlinear effects are expected to be present and hence 
it is  interesting to investigate their influence.

\begin{figure}[ht!]
\begin{center}
\subfigure[ ]{\label{fig:first1}
\includegraphics[width=0.488\textwidth]{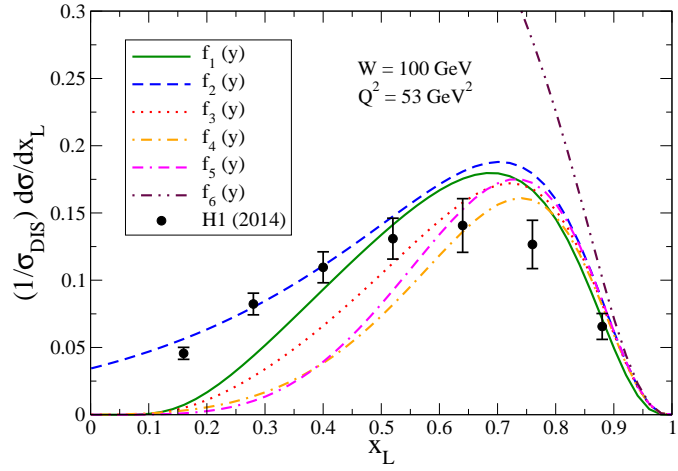}}\\
\subfigure[ ]{\label{fig:second1}
\includegraphics[width=0.488\textwidth]{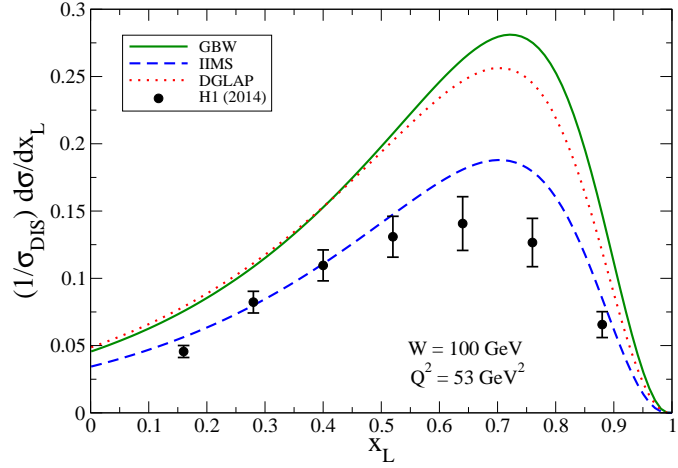}}\\
\subfigure[ ]{\label{fig:second1}
\includegraphics[width=0.488\textwidth]{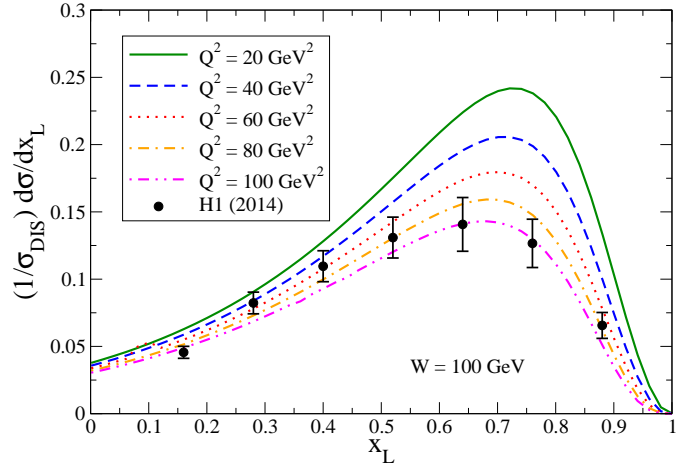}}
\end{center}
\caption{Leading neutron spectra compared with experimental data at $W=100$ GeV considering 
(a) the IIMS scattering amplitude  and different choices of the pion flux ($Q^2= 53$ GeV$^2$);
(b) the  $f_2$ pion flux and different choices for the scattering amplitude ($Q^2= 53$ GeV$^2$); and  
(c) the $f_2$ pion flux and the IIMS scattering amplitude for  different values of photon virtuality $Q^2$.}
\label{fig4}
\end{figure}

We address now the dependence of our predictions for the leading neutron spectra on the models of the pion flux and of the dipole-pion scattering amplitude.
 It is important to emphasize that the spectrum is proportional to the product of these two quantities. Consequently, in what follows we will initially 
assume a given model for one of these quantities and analyse the dependence on the other quantity. Having done that, we will choose one combination, calculate 
the normalized cross section $(1/\sigma_{DIS}) d \sigma / d x_L$ and compare  with the experimental data. The normalization was taken from Ref. \cite{snorm} and  
is given by:
\begin{equation} 
\sigma_{DIS} = \frac{4 \pi^ 2 \alpha_{em}}{Q^ 2} \frac{c}{x^{\beta}}
\label{signorm}
\end{equation}
where $c=0.18$, $\beta = d \, \ln (Q^2 / \Lambda_0^2)$ with $d = 0.0481$ and $\Lambda_0 = 0.292$ GeV. 
In Fig.  \ref{fig4} (a) we  consider the IIMS model for the scattering amplitude and estimate the leading neutron spectrum using the different 
models of the pion flux discussed in the previous section. We observe that  the behavior at medium and small values of $x_L$ is strongly dependent 
on the choice of the model. Using the  cut-off $\Lambda_c$ in the range defined in Ref. \cite{wally2},  our results suggest that  model $f_6$ is 
disfavored. In Fig.  \ref{fig4} (b) we consider the $f_2$ model for the pion flux and estimate the spectra using different models of the scattering amplitude.
 We can see that the predictions have similar behavior in
the transition between the small and large-$x_L$ regimes, but with the magnitude being dependent on the description of the QCD dynamics. In the 
previous analysis we have considered $Q^2 = 53$ GeV$^2$. However, the experimental data were obtained for a wide range of virtualities, which 
implies that there is some arbitrariness present in the choice of the value of $Q^2$ used in our calculations. In order to estimate the 
uncertainty present in this choice, in Fig. \ref{fig4} (c) we consider the $f_2$ and IIMS models and calculate the spectrum for different 
values of the photon virtuality. We observe that our predictions are more compatible with the experimental data if larger values of $Q^2$ are assumed.

So far we have assumed that the photon hits only the pion, in a type of ``impulse aproximation''. However it has been shown in \cite{pirner} 
that very often the photon hits 
also the neutron, specially in the low $Q^2$ domain, where the photon has low resolving power. In these cases the extra interactions generate
 the so called absorptive corrections,  which can be estimated with models. In the case of leading neutron production, it has been shown 
\cite{pirner} that the corrections 
are not very large and affect almost uniformly all the $x_L$ spectrum. Using the results of \cite{pirner}, we shall here simply multiply 
our spectra by the absorptive correction factor, $K_{abs} = 0.7$.

\begin{figure}
\centerline{\psfig{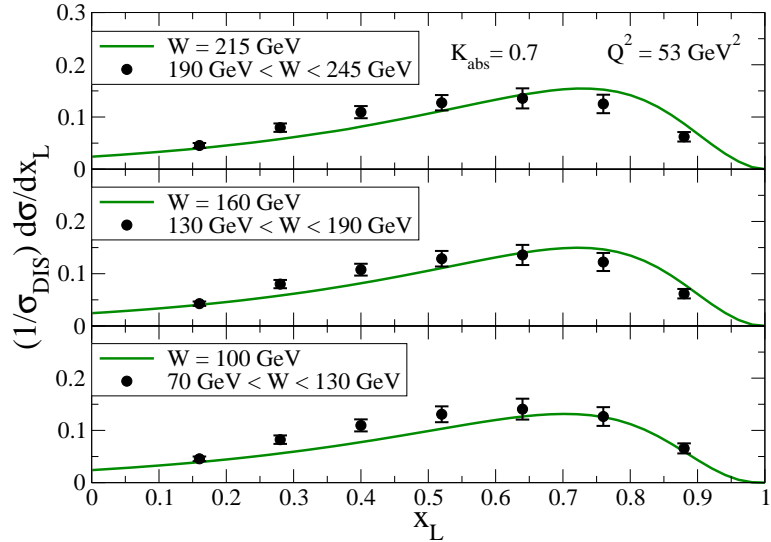}}
\vskip0.5cm
\caption{Leading neutron  $x_L$ spectrum  for three photon-proton energies compared with recent H1 data. }
\label{fig6}
\end{figure}

In Fig. \ref{fig6} we give our description of data, which is quite reasonable. 
All the parameters contained in the dipole cross section have been already fixed by the analysis of other DIS data from HERA. The pion flux 
has been fixed by fitting data of low energy hadronic collisions.  
As metioned above, it has been argued that the factor $\frac{2}{3}$ predicted 
by the additive quark model to relate the photon-pion and  photon-proton cross sections   
could  be smaller, closer to 
$\frac{1}{3}$. Using this different factor, we would then underpredict 
the data in the low $x_L$ region. In principle, this would be acceptable since in this 
region other processes are expected to play  a  role. 
As already emphasized, this subject deserves more detailed studies.

Finally, we  analyse the Feynman scaling in the leading neutron spectra and the contribution of nonlinear effects for this process.
 As mentioned before this process is surprisingly sensitive to low $x$ physics. In particular, the large $x_L$ ($x_L > 0.3$) we observe 
the transition from the large to small $x$ domain. It is therefore interesting to check whether gluon saturation effects are already 
playing a significant role. As it can be seen from Eqs.  (\ref{crossgen}) and ({\ref{f3}), all the energy dependence is contained in 
the dipole forward scattering amplitude $\cal{N}$.  Therefore, we can  estimate  the contribution of the nonlinear effects 
 comparing the results obtained with saturation model, e.g. the IIMS model given in (\ref{CGCfit}), with  those obtained with the linear
 model, Eq. (\ref{sig_dglap}).  Moreover, the theoretical expectation can be obtained using the GBW model for the scattering amplitude, 
Eq. (\ref{gbw}). In the linear limit, when the dipole radius is very small (or equivalently 
$Q^2$ is very large) or the saturation scale is very small (and hence the energy is not very high), we can expand the exponent and obtain 
\beq
\sigma_{d\pi} (r,\hat{x})  \propto \sigma_0 \, {\cal{N}}(r,\hat{x}) \simeq \sigma_0  \, \frac{Q^2_s(\hat{x}) r^2}{4}   
\simeq   \sigma_0 \,   Q^2_0 \, {x_0}^{\lambda} \left[\frac{(1-x_L)W^2 + Q^2}{Q^2 + m^2_f}\right]^{\lambda} \,\,.
\label{aproxlin}
\eeq
Consequently, in this regime we see that the leading neutron $x_L$ spectrum will depend on $W$. In a complementary way, in the nonlinear 
limit,  when the dipole radius is very large 
(or equivalently  $Q^2$ is very small) or the saturation scale is very large (and hence the energy is  very high), we obtain 
\beq
\sigma_{d\pi} (r,\hat{x})  \propto \sigma_0 \, {\cal{N}}(r,\hat{x}) \simeq \sigma_0  \
\label{aproxnonlin}
\eeq
which is energy independent.  One could argue that the mere inspection of  Eq. (\ref{crossgen}) would suggest that at some asymptotically 
high energy the 
photon-pion cross section would reach some ``black disk'' limit and the energy dependence would disappear. We would like to emphasize that
 the information contained 
in (\ref{aproxlin}) and (\ref{aproxnonlin}) is much richer and indicates the route through which the asymptotic limit is reached and the
 role played by nonlinear 
effects.  These expectations can be compared with those obtained using the IIMS and DGLAP models for the dipole-pion cross section. 
In Fig. \ref{fig8} (a) we show the spectra obtained in a purely  linear approach. As expected we see a noticeable energy dependence.  
In contrast, the nonlinear predictions presented  in Fig. \ref{fig8} (b) show  a remarkable suppression of the  energy dependence at 
low values of $Q^2$, consistent with the expectations. These results indicate that the Feynman scaling (and how it is violated) can be directly
related to the QCD dynamics at small-$x$.

\begin{figure}[t]
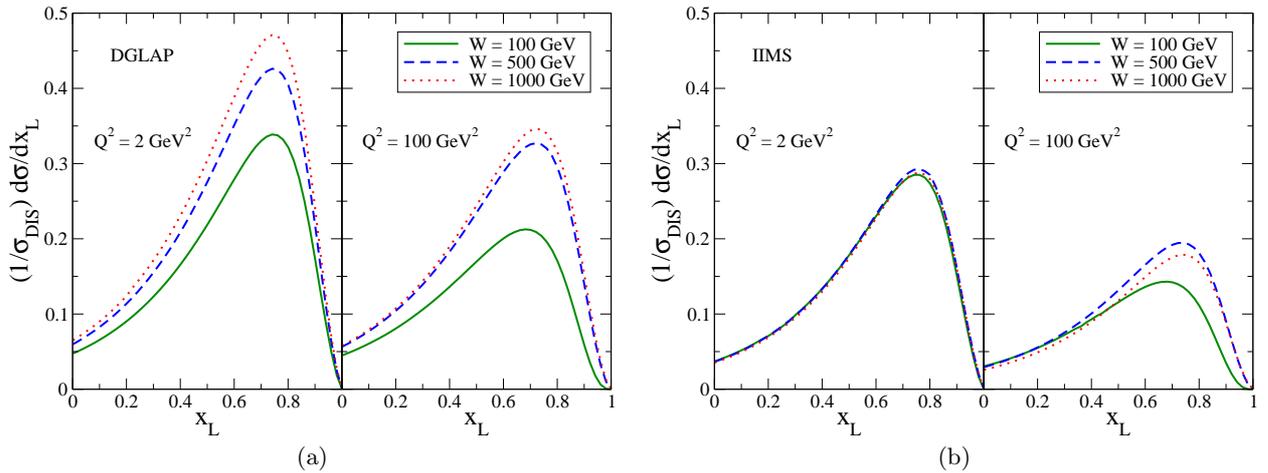

\vskip0.3cm
\begin{tabular}{ccc} 
\psfig{figure=leadf9n.eps,width=8cm}& \,\,\, & \psfig{figure=leadf10n.eps,width=8cm}\\
(a) & \,\,\, &(b)
\end{tabular}
\caption{(a) Leading neutron spectra for different energies considering 
(a) the linear DGLAP model  and (b) the nonlinear IIMS model  for the dipole-pion cross section.}
\label{fig8}
\end{figure}


\section{Summary}
\label{conc}

In this work we have studied leading neutron production in  $e + p \rightarrow e + n + X$ collisions at high energies 
and have calculated the Feynman $x_L$ distribution of these neutrons. The  differential cross section was written  in terms 
of the pion flux and the  photon-pion total cross section. We have proposed to describe this process using the color dipole formalism and, assuming the validity of the additive quark model, we have related the dipole-pion with the well determined dipole-proton cross section. In this formalism we have been able to estimate the dependence of the predictions on the description of the QCD dynamics at high energies as well as the contribution of the gluon saturation effects for the leading neutron production. With the parameters strongly constrained by other 
phenomenological information, we were able to reproduce the recently released H1 leading neutron spectra. 
One of our most interesting conclusions is that leading neutron spectra can be used to probe the low $x$ content of the pion target and hence it is a new observable where we can look for gluon saturation effects. At higher energies this statement will become more valid. Moreover saturation physics provides a precise route to Feynman scaling, which will eventually happen at higher energies.  These conclusions  motivate more detailed studies, in particular in the subjects which are the main source of theoretical uncertainties in our calculations: i) the validity of the additive quark model; ii) the factor $K_{abs}$ accounting for the absorptive corrections, which are model dependent; iii) the precise form of the pion flux; iv) the precise form of  the dipole cross section.  It is important to emphasize that none of these numbers or functions is free. On the contrary, they are subject to severe constraints from other experimental information and the freedom to choose them will be further reduced in the future.

\begin{acknowledgments}

We are deeply grateful to W. Melnitchouk and to C.R. Ji for enlightening discussions. 
This work was  partially financed by the Brazilian funding agencies CNPq, CAPES, FAPERGS and FAPESP.

\end{acknowledgments}

\hspace{1.0cm}

\end{document}